\documentclass{eas}
\usepackage{graphicx}
\usepackage{amsmath}
\begin{document}

\title{Linear approximation of seismic inversions:
new kernels and structural effects.} 
\runningtitle{Linear approximation of seismic inversions}
\author{G. Buldgen}\address{Institut d'Astrophysique et G\' eophysique, University of Li\`egge, Li\`ege, Belgium}
\author{D. R. Reese}\address{LESIA, Observatoire de Paris.}
\author{M. A. Dupret}\address{Institut d'Astrophysique et G\' eophysique, University of Li\`ege, Li\`egge, Belgium }
\begin{abstract}
Thanks to the space-based photometry missions CoRoT and Kepler, we now benefit from a wealth of seismic data for stars other than the sun. In the future, K2, Tess and Plato will provide further observations. The quality of this data may allow kernel-based linear structural inversion techniques to be used for stars other than the sun.
To understand the limitations of this approach, we analyse the validity of the linear assumption used in these inversion techniques. We inspect various structural pairs and see how they are affected by structural changes. 
We show that uncertainties in radius strongly affect structural pairs of nondimensional variables, and that various other effects might come into play. Amongst these, the importance of micro-physics give the most striking example of how uncertainties in stellar models impact the verification of the linear relations. We also point out that including seismic constraints in the forward modelling fit helps with satisfying the linear relations.
\end{abstract}
\maketitle
\section{Introduction}
Asteroseismology is considered the golden path to study stellar structure. This research field benefits from high quality data for a large sample of stars stemming from the successes of the CoRoT, Kepler and K2 missions. The detection of solar-like oscillations in numerous stars allows a more accurate study of stellar structure. In the future, the Tess and Plato missions will carry on this space-photometry revolution. The successes of asteroseismology were preceded by those of helioseismology, the study of solar pulsations. Indeed, the quality of seismic data of the sun is still far beyond what is achievable for other stars, even in the era of the space missions. In helioseismology, numerous methods were developed to obtain constraints on the solar structure. Amongst them, inversion techniques lead to the determination of the base of the convective envelope, the helium abundance in this region and the rotational profile of the sun. The determination of the sound speed and density profiles also demonstrated the importance of microscopic diffusion for solar models.
\\
\\
In asteroseismology, the use of these inversion techniques can now be considered for a limited number of targets, under the conditions of validity of all the hypotheses hiding behind the basic equations defining their applicability domain. The most constraining of these hypotheses is to assume a linear relation between frequency differences and structural differences. In this paper, we present some issues surrounding the linearity of these relations for various structural pairs and more specifically for kernels of the convective parameter.
\vspace{-0.2cm}
\section{Obtaining new kernels}
\subsection{Theoretical developments}
The inversion procedure relies on the variational principle and the frequency-structure relation for adiabatic stellar oscillations. The fundamental equations with which are carried out the inversions are the following: 
\begin{align}
 \frac{\delta \nu_{n,\ell}}{\nu_{n,\ell}}=\int_{0}^{1} K^{n,\ell}_{s_{1},s_{2}}\frac{\delta s_{1}}{s_{1}}dx + \int_{0}^{1} K^{n,\ell}_{s_{2},s_{1}}\frac{\delta s_{2}}{s_{2}}dx \label{EqVar}
  \end{align}
with: $\frac{\delta s}{s}=\frac{s_{obs}-s_{ref}}{s_{ref}}$, where $s_{1}$, $s_{2}$ are structural variables such as $\Gamma_{1}$, $c^{2}$, $\rho_{0}$, ... $K^{n,l}_{s_{i},s_{j}}$ the structural kernels. Since we work with models, we have dropped the surface correction terms. 

Changing the kernels is crucial in asteroseismology, since we don't have as many frequencies as in helioseismology. In practice, this means finding the functions $K^{n,l}_{s_{3},s_{4}}$ and $K^{n,l}_{s_{4},s_{3}}$ associated with the structural pair $(s_{3},s_{4})$ satisfying the following relation:
\begin{align}
 \frac{\delta \nu_{n,\ell}}{\nu_{n,\ell}}=\int_{0}^{1} K^{n,\ell}_{s_{1},s_{2}}\frac{\delta s_{1}}{s_{1}}dx + \int_{0}^{1} K^{n,\ell}_{s_{2},s_{1}}\frac{\delta s_{2}}{s_{2}}dx =\int_{0}^{1} K^{n,\ell}_{s_{3},s_{4}}\frac{\delta s_{3}}{s_{3}}dx + \int_{0}^{1} K^{n,\ell}_{s_{4},s_{3}}\frac{\delta s_{4}}{s_{4}}dx \nonumber
\end{align}
This is done using either the method of conjugated functions (\cite{Ell}, \cite{Kos}) or the `direct' method (\cite{Bul}).

Both methods have advantages and inconvenients, we list them below:
\vspace{-0.3cm}
\begin{itemize}
\item \textbf{Method of conjugated functions:} system of $1^{st}$ order equations, involving a vector function related to the structural kernels. Simple coefficients but requires knowing of mass.
\item \textbf{Direct method:} $2^{nd}$ to $3^{rd}$ order differential equation for the kernels. No need to know of mass or radius to define boundary conditions.
\end{itemize}
\subsection{Examples of new structural kernels}
\begin{figure*}[t]
	\centering
		\includegraphics[width=8.7cm]{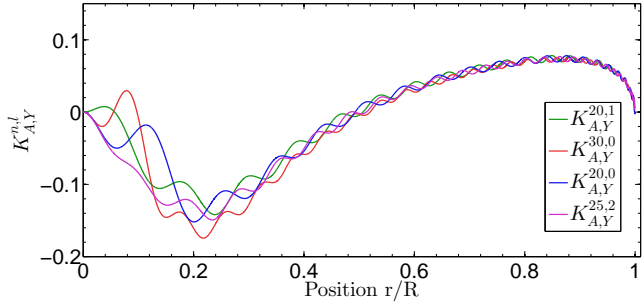}
	\caption{Kernel associated with the convective parameter, A, in the $(A,Y)$ structural pair.}
		\label{figVar}
\end{figure*}

\begin{figure*}[t]
	\centering
		\includegraphics[width=8.7cm]{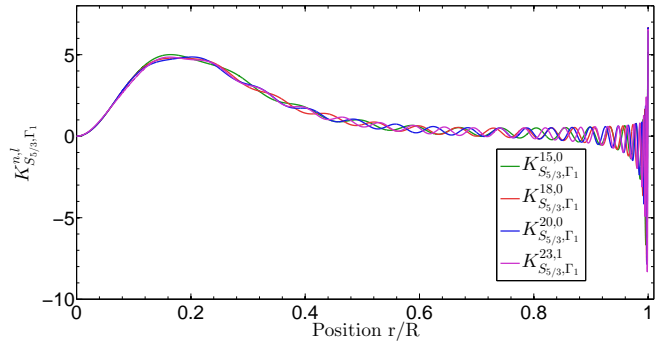}
	\caption{Kernel associated with the entropy proxy, $S_{5/3}$, in the $(S_{5/3},Y)$ structural pair.}
		\label{figKerS}
\end{figure*}

In this section, we illustrate structural kernels for two new pairs, the $(A,Y)$ pair, and the $(S_{5/3},Y)$ pair, which have both important characteristics for helio- and asteroseismology. Examples of kernels associated with $A=\frac{1}{\Gamma_{1}}\frac{d\ln P}{d\ln r}-\frac{d \ln \rho}{d\ln r}$, the convective parameter, are shown in Fig. \ref{figVar} and examples of kernels associated with the entropy proxy, $S_{5/3}=\frac{P}{\rho^{5/3}}$ are shown in Fig. \ref{figKerS}.
One can easily obtain the following structural pairs: $(P,\Gamma_{1})$, $(u,\Gamma_{1})$, $(c^{2},\Gamma_{1})$, $(g,\Gamma_{1})$, $(S_{5/3},\Gamma_{1})$, $(\frac{P}{\rho^{\Gamma_{1}}},\Gamma_{1})$, $(A,\Gamma_{1})$, + all of these with $Y$ as secondary variable instead of $\Gamma_{1}$. 
\section{Illustration of non-linearities}

Investigating the impact of non-linearities is extremely important since in the context of asteroseismology, they may well be a limitation to the application of inversion techniques. In which case, analysing the sources of these non-linearities may point towards improvements of stellar models but also further developments of adapted seismic analysis techniques. To test their impact, we build a target model with a given set of physics (chemical abundances, $\alpha_{MLT}$, mixing, ...) and fit this artificial target with a reference model of the same mass. To ensure that both target and reference models have the same radius, we use the age as a free parameter to fit the mean density of the target model. We can also analyse effects of errors on the radius by fitting very accurately a slightly modified value of the mean density of the target model. By not adding seismic constraints, we ensure that the reference model is \textit{at the limit of non-linearity}.
\begin{figure*}[t]
	\centering
		\includegraphics[width=8.6cm]{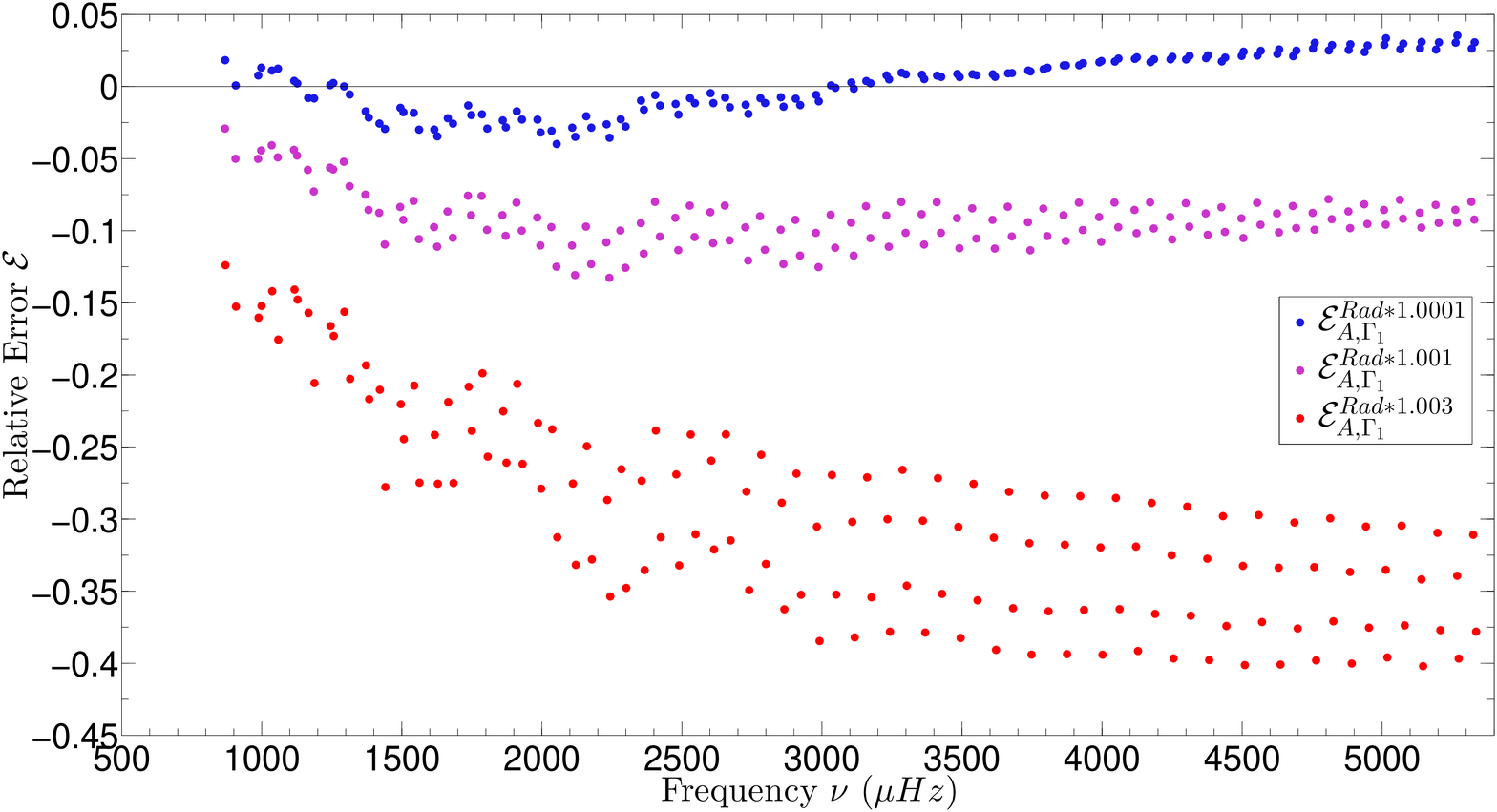}
	\caption{Effects of radius changes on the verification of Eq. \ref{EqVar} for various structural pairs.}
		\label{figVarRad}
\end{figure*}
\begin{figure*}[t]
	\centering
		\includegraphics[width=8.6cm]{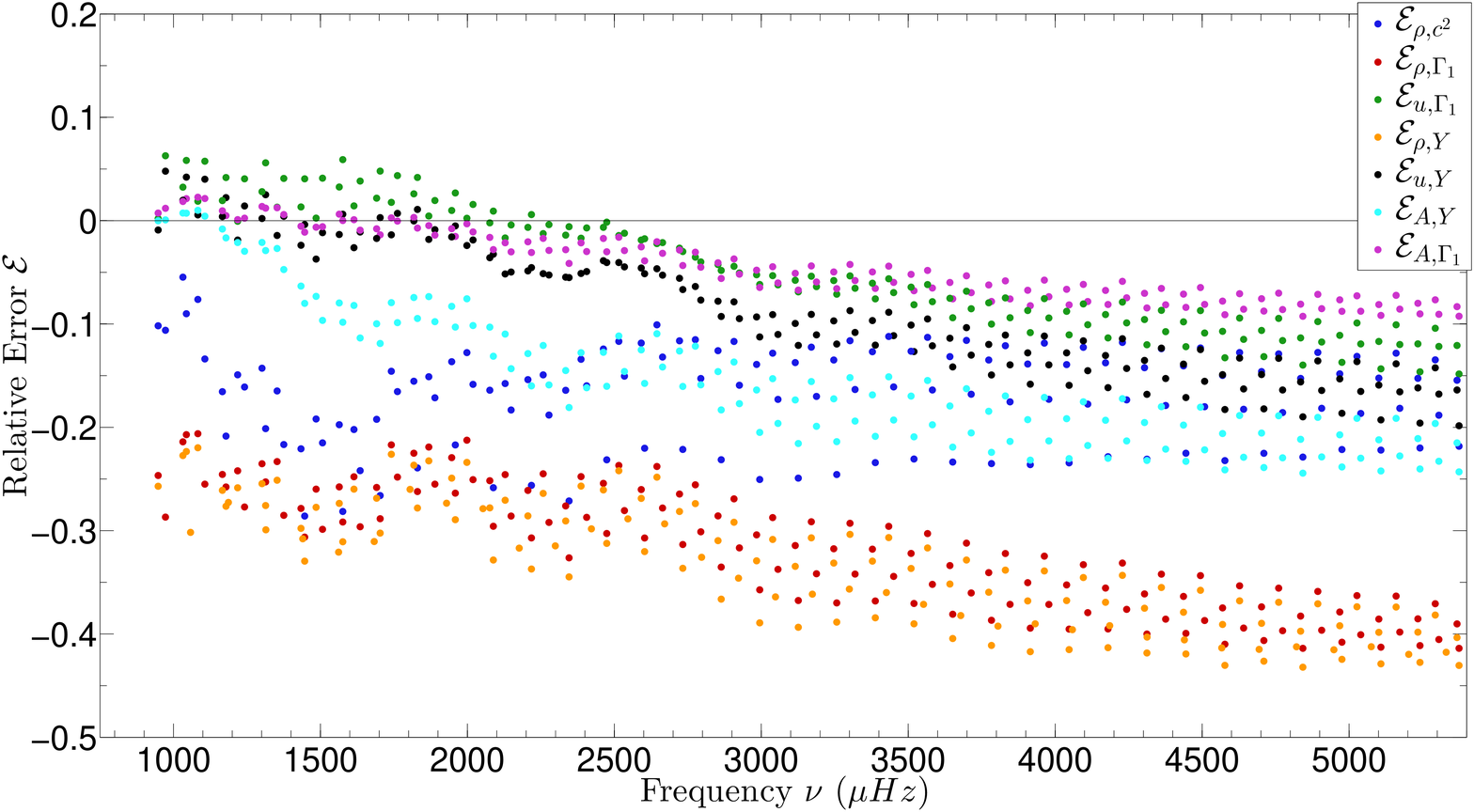}
	\caption{Effects of the opacities on the verification of Eq. \ref{EqVar} for various structural pairs.}
		\label{figVarGN}
\end{figure*}
From Fig. \ref{figVarRad}, we see that for the $(A,Y)$ and $(A,\Gamma_{1})$ pairs, the knowledge of the radius is crucial to ensure a sufficient verification of the linear relation. 

In fig \ref{figVarGN}, we illustrate effects of a change in the heavy element mixture (keeping $Z$ constant), thus mimicking a strong change in opacity. This induces strong non-linear effects between the models.
\vspace{-0.4cm}
\section{Effects of metallicity}
\begin{figure*}[t]
	\centering
		\includegraphics[width=8.1cm]{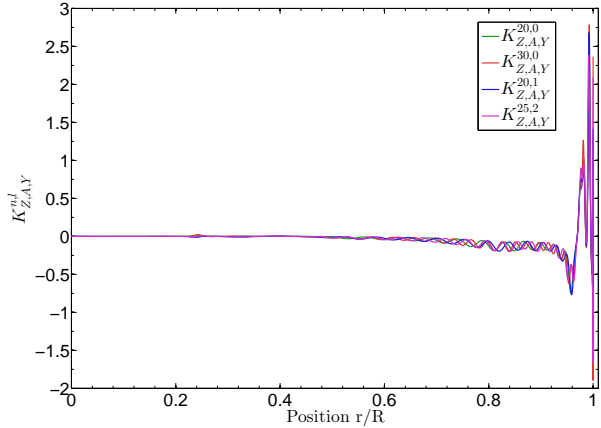}
	\caption{Kernels associated with the metallicity, $Z$, in the $(A,Y,Z)$ structural triplet.}
		\label{figVarZ}
\end{figure*}

The impact of the metallicity in the linear development of the linear perturbations of $\Gamma_{1}$ is always neglected. While the change in metallicity will be small, in some cases the impact on thermal structure is sufficient to induce non-linear effects, which are seen with the $(A,Y,Z)$ structural triplet. This is particularly the case if the differences of $Y$ are very small. We illustrate in Fig. \ref{figVarZ} the kernels associated with $Z$ in the $(A,Y,Z)$ triplet. One can see that their amplitude is comparable to that of helium kernels, meaning that $Z$ can impact directly the linear relations, but also indirectly through its influence on the temperature gradient in the model. 
\vspace{-0.8cm}
\section{Prospects and conclusions}
The new kernels have allowed us to explore in depth the verification of the linear relations between frequencies and structural quantities. These kernels also allow us to apply inversions of various integrated quantities to fully exploit the data of the Kepler and CoRoT missions. The verification of the linear relations can vary from one structural pair to another at the limit of non-linearity. Using multiple pairs can ensure the robustness of the inverted results.

New metallicity kernels and kernels involving a proxy for entropy, $S_{5/3} = P/\rho^{5/3}$, can be used to re-analyse the solar metallicity problem. The $(S_{5/3},Y)$ kernels can also be used in the asteroseismic context for integrated quantities inversions.

\end{document}